\documentclass[aps,prl,twocolumn,aps,floatfix]{revtex4-1}

\usepackage{graphicx}
\usepackage{dcolumn}
\usepackage{bm}
\usepackage{amsmath, amssymb, amsfonts}

\newcommand{\omz}{\omega_z}
\newcommand{\omc}{\omega_c}
\newcommand{\oma}{\omega_a}

\newcommand{\omr}{\omega_r}
\newcommand{\zhat}{\hat{z}}
\newcommand{\nhat}{\hat{n}}
\newcommand{\nbar}{\bar{n}}
\newcommand{\nmax}{\bar{n}_{\textrm{max}}}
\newcommand{\Dca}{\Delta_{\textrm{ca}}}
\newcommand{\Dpc}{\Delta_{\textrm{pc}}}
\newcommand{\DNzero}{\Delta_{N}^{(0)}}
\newcommand{\hamiom}{\mathcal{H}}
\newcommand{\Zcm}{Z_{\mbox{\scriptsize{cm}}}}

\begin{document}
\title{Tunable Cavity Optomechanics with Ultracold Atoms}
\author{T.P.\ Purdy$^1$}
\author{D.W.C.\ Brooks$^1$}
\author{T.\ Botter$^1$}
\author{N.\ Brahms$^1$}
\author{Z.-Y.\ Ma$^{1}$}
\thanks{Present address: Shanghai Institute of
Optics and Fine Mechanics, Chinese Academy of Sciences.}
\author{D.M.\
Stamper-Kurn$^{1,2}$} \email{dmsk@berkeley.edu}
\affiliation{
    $^1$Department of Physics, University of California, Berkeley CA 94720, USA \\
    $^2$Materials Sciences Division, Lawrence Berkeley National Laboratory, Berkeley, CA 94720, USA}
\date{\today}%

\begin{abstract}
We present an atom-chip-based realization of quantum cavity
optomechanics with cold atoms localized within a Fabry-Perot cavity.
Effective sub-wavelength positioning of the atomic ensemble allows
for tuning the linear and quadratic optomechanical coupling
parameters, varying the sensitivity to the displacement and strain
of a compressible gaseous cantilever. We observe effects of such
tuning on cavity optical nonlinearity and optomechanical frequency
shifts, providing their first characterization in the
quadratic-coupling regime.
\end{abstract} \maketitle

Experimental realizations of cavity optomechanics, wherein the
motion of a mechanically compliant cantilever is measured by its
interaction with an electromagnetic resonator, serve as paradigms
for understanding open quantum systems and the limits of quantum
measurement \cite{kipp08sciencereview}. In most realizations, cavity
optomechanical phenomena, such as optical cooling
\cite{brag67,kipp05anal,arci06,giga06cooling} and confinement
\cite{shea04spring,corb07gram} of the cantilever or optical
nonlinearity \cite{dors83bistability,gupt07nonlinear} and squeezing
\cite{manc94,fabr94noise}, arise from the dominant linear coupling
between cavity photons and the cantilever position.  Recent
experiments on thin SiN membranes positioned within a Fabry-Perot
cavity \cite{thom08membrane,sank10tunable} have highlighted the new
capabilities afforded by quadratic optomechanical coupling, such as
measurement of the cantilever energy and of phonon shot noise
\cite{cler10shotnoise}.

Here, we demonstrate a tunable cavity optomechanical system
constructed by integrating a microfabricated atom chip with a
Fabry-Perot optical resonator.  We tune the optomechanical
sensitivity to the displacement and the strain of an atomic
ensemble, arising from linear and quadratic optomechanical coupling,
respectively, by positioning cold atoms with nm-scale precision
within the resonator mode. The ensemble thereby serves as the
quantum analogue of the SiN membranes used in recent experiments
\cite{thom08membrane,sank10tunable}.  We study effects of tunable
coupling on optomechanical bistability and the optomechanical
frequency shift, providing the first characterization of
optomechanical effects in the quadratic-coupling regime.  The
agreement between our measurements and theory establishes the
equivalence of cavity optomechanical systems using either solid- or
gas-phase cantilevers.

We begin by adapting recent theoretical descriptions of cavity
optomechanics using cold atoms \cite{murc08backaction} to highlight
the new capabilities presented in this work.  Consider the motion of
$N$ identical atoms along the axis ($\zhat$) of a Fabry-Perot
optical resonator and confined within a harmonic potential with
mechanical frequency $\omz$ and centered at position $z_0$.  The
interaction of atom $i$ at $z_i = z_0 + \delta z_i$ with the cavity
field is characterized by the angular frequency $g(z_i) = g_0
\sin(\phi_0 + k_p \delta z_i)$, where $\phi_0 = k_p z_0$, $k_p$ is
the resonant cavity wavevector and $g_0$ is determined by the cavity
mode volume, the optical frequency, and atomic dipole matrix
elements. Assuming the detuning $\Dca = \omc - \oma$ between the
cavity ($\omc$) and the atomic-electronic ($\oma$) resonance
frequencies is large, we retain the dispersive atoms-cavity
interaction, and expand to second order in the Lamb-Dicke parameters
$k_p \delta z_i \ll 1$, obtaining the following interaction
Hamiltonian:
\begin{equation}
\hamiom \! \simeq\!  \left[\hbar \DNzero \!- \!F \Zcm \sin 2 \phi_0
\!-\! F k_p \left( \Zcm^2 + \sigma^2 \right) \cos 2 \phi_0 \right]\!
\nhat. \nonumber
\end{equation}
Here, the zero-order term defines the static dispersive shift of the
cavity resonance, $\DNzero = N (g_0^2/\Dca) \sin^2 \phi_0$, with all
atoms held at the trap center. The next term describes linear
optomechanical coupling to the ensemble center of mass $\Zcm$, acted
upon by the per-photon force defined as $F = N \hbar k_p g_0^2 /
\Dca$. The second-order term describes quadratic optomechanical
coupling. We identify both a quadratic sensitivity to the cantilever
mode and also a coupling to the position variance $\sigma^2$ of the
atomic medium. The relative strength of the linear and quadratic
couplings is controlled by the position of the ensemble along the
cavity axis.

\begin{figure}[tb]
\begin{center}
\includegraphics[angle = 0, width =.4\textwidth]{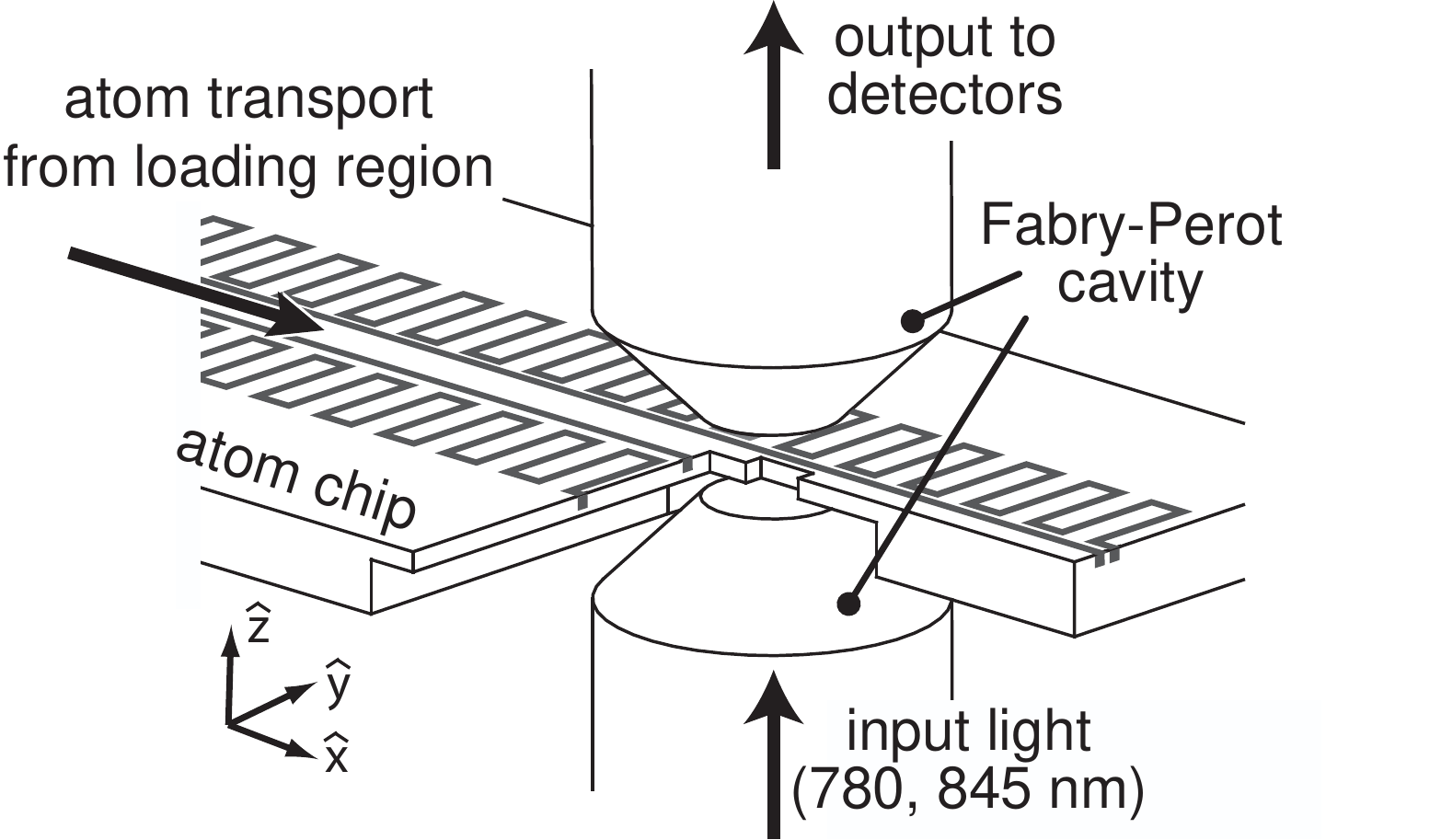}
\end{center}
\caption{An atom chip is used to transport and position atoms within
a Fabry-Perot cavity, where they act as a compressible cantilever
for studies of cavity optomechanics. A cutaway cross section shows
the profile of the chip, with copper wires shown in gray.}
\label{fig:schematic}
\end{figure}

In previous experiments \cite{gupt07nonlinear,bren08opto}, the broad
spatial extent of the atomic ensemble suppressed the quadratic
optomechanical coupling, while maintaining a linear coupling to
phonon-like perturbations of the trapped gas. Here, we overcome this
limitation by localizing the gas within a sub-micron region
positioned variably along the cavity axis. Such localization is
achieved using a combination of magnetic and optical traps produced
using an atom chip (Fig.\ \ref{fig:schematic}).  This chip is
fabricated by embedding 75-$\mu\mbox{m}$-thick electroplated copper
electromagnet wires into a $500\, \mu\mbox{m}$ thick silicon
substrate. These electromagnets are used to trap $^{87}$Rb atoms and
to transport them 2.2 cm along the chip surface into the mode-volume
of a Fabry-Perot optical resonator, formed by curved mirrors placed
above and below the chip surface.  At the location of the cavity,
the substrate is thinned to $100 \, \mu\mbox{m}$, allowing the
mirrors to be spaced by $250 \, \mu\mbox{m}$, and a $200 \times 400
\, \mu\mbox{m}$ rectangular hole is etched through the chip to allow
the cavity field to propagate unobstructed.

The cavity, with Gaussian mode waist of 25 $\mu$m, is stabilized
with one of its TEM$_{00}$ resonances near the $^{87}$Rb D2
transition, at wavenumber $k_p = 2 \pi / (780 \,\mbox{nm})$. For
this mode, the cavity finesse of $1.7 \times 10^5$ results from
transmissions of 1.5 and 12 ppm through the input and output
mirrors, respectively, and additional losses of 26 ppm.  An atom at
the antinode of the cavity field, driven at the atomic resonance,
experiences a coupling of $g_0\!=\!2\pi\!\times\!13.1$ MHz, giving a
single-atom cooperativity of $g_0^2/2\kappa\Gamma=16$ where
$\kappa\!=\! 2\pi \!\times\! 1.8$ MHz and $\Gamma\!=\!=2\pi
\!\times\! 3$ MHz are the cavity and the atomic half line-widths,
respectively.

Once within the cavity, the magnetically trapped gas, with rms width
less than 400 nm along the cavity axis, is transferred into the
wells of a one-dimensional optical lattice trap formed by light with
wavenumber $k_t = 2 \pi / (845 \,\mbox{nm})$ driving a second
TEM$_{00}$ cavity mode \cite{murc08backaction,colo07}. This transfer
is accomplished by ramping the optical lattice trap depth to 10
$\mu$K before decompressing and switching off the magnetic trap. The
lattice depth is lowered to 1.5 $\mu$K to cool the atomic gas
further, and then raised to provide a tunable axial oscillation
frequency. The wells of this lattice trap serve as the near-harmonic
mechanical potential for the several-thousand-atom ensembles used in
our realization of cavity optomechanics. Time-of-flight temperature
measurements on atoms released from the lattice imply an axial
ground state occupation of $>90 \%$.  Thus, our system realizes
tunable cavity optomechanics in the quantum regime.

\begin{figure}[tb]
\begin{center}
\includegraphics[angle = 0, width =.4\textwidth]{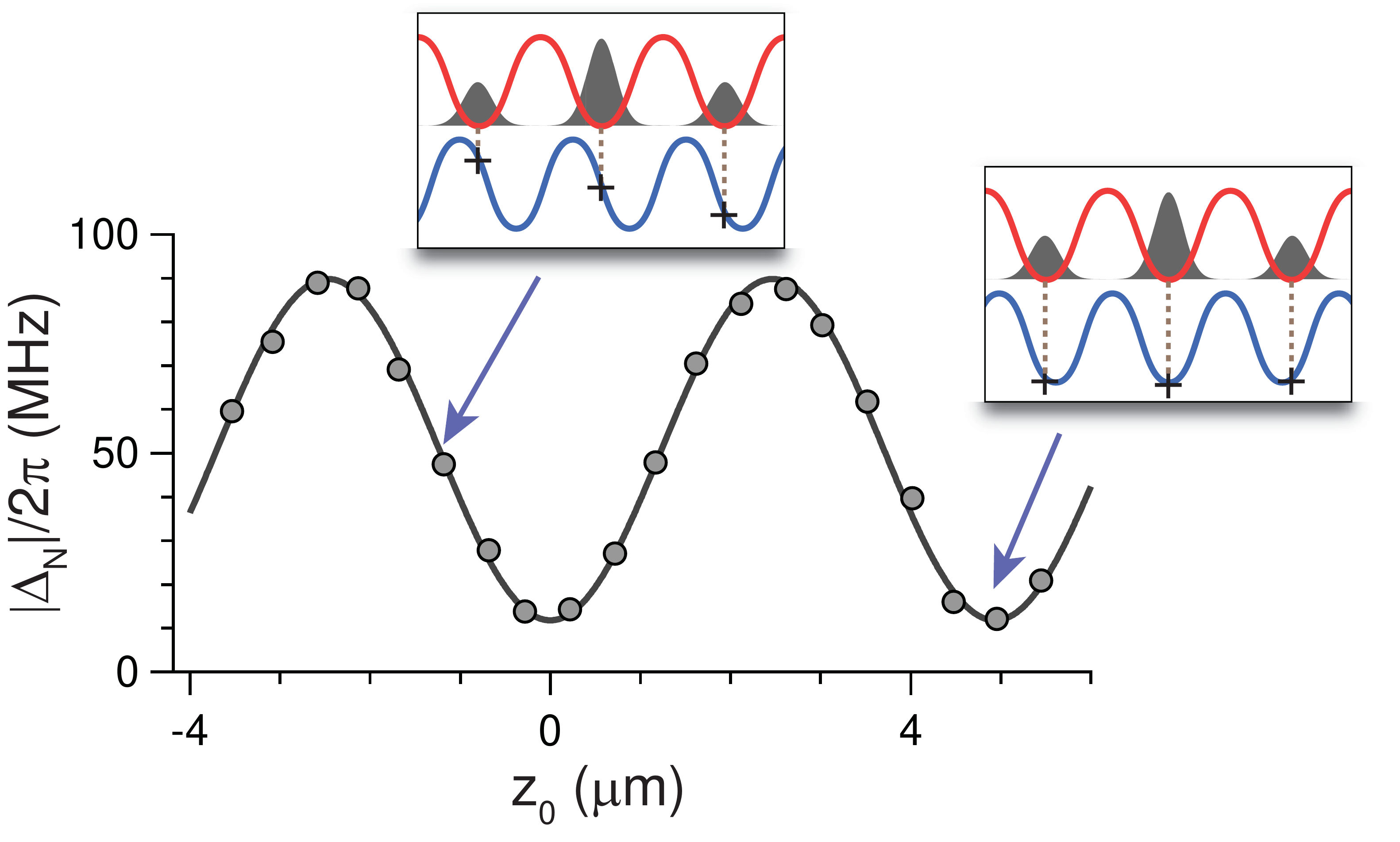}
\end{center}
\caption{(Color) The atom-induced shift of the cavity resonance at
low probe power varies sinusoidally (fit shows contrast of 0.77)
with the central atomic position $z_0$. Panels show estimated atomic
distributions (gray), relative to the lattice trap potential (red,
top) and cavity probe intensity (blue, bottom), at maximum linear
(left) or quadratic (right) coupling. Here $N\!=\!3500$, $\Dca/2\pi
=-6$ GHz, $\omz/2\pi \!=\! 70$ kHz.} \label{fig:deltan_vs_pos}
\end{figure}

The position of the ensemble within the resonator is visible in the
atom-induced shift of the cavity resonance, measured via the
transmission of a low-power probe beam swept across the cavity
resonance (Fig.\ \ref{fig:deltan_vs_pos}).   As expected, this
frequency shift oscillates sinusoidally with the ensemble position
with a spatial period $\pi/ (k_p-k_t) = 5 \, \mu\mbox{m}$.

The contrast of this oscillation provides information on the
distribution of atoms within the intracavity optical lattice.
Assuming the axial density distribution of atoms within a lattice
site to be Gaussian with rms width $\sigma=\sqrt{\hbar / 2 m \omz}$,
atoms in a single lattice site shift the cavity resonance by
\cite{note:transverse}
\begin{equation}
\Delta_N = N \frac{g_0^2}{2 \Dca} \left( 1 - e^{-2 k_p^2 \sigma^2}
\cos 2 \phi_0 \right). \label{eq:shiftwithsigma}
\end{equation}
The measured contrast is lower than this single-site limit,
indicating that atoms are distributed among neighboring wells of the
optical lattice.  Assuming a Gaussian distribution of atoms among
wells reduces the contrast by $e^{-2 (k_p - k_t)^2 \Sigma^2}$, we
estimate the atoms to be divided among wells spanning $\Sigma
\lesssim 400 \,\mbox{nm}$ along the cavity axis; that is, the
ensemble is not confined to a single lattice site, though the
majority of atoms occupy no more than two sites. We nevertheless
achieve fine control of the optomechanical coupling owing to the
near-equivalence of neighboring lattice sites, between which
$\phi_0$ varies only by 0.26 rad. Thus, by using two similar
wavelengths for the optical trapping and the probe light we parlay
the moderate confinement of our chip-based magnetic trap to achieve
sub-wavelength control of the gas-phase cantilever.

We use this tunable system to study the influence of variable linear
and quadratic coupling parameters on two optomechanical effects.  We
consider first the cavity-optical nonlinearity that arises from
steady-state modification of the atomic cantilever in response to
the optical potential from an average of $\nbar$ probe photons
within the cavity.  This potential both displaces the center of mass
and also changes the rms width of the compressible ensemble. These
variations, ascribed to the linear and the quadratic optomechanical
coupling, respectively, affect the atom-induced shift
$\Delta_N(\nbar)$ of the optical cavity resonance, now a function of
$\nbar$, according to Eq.\ \ref{eq:shiftwithsigma}. The value of
$\nbar$ is determined self-consistently from the Lorentzian cavity
response function, $\bar{n}=\nmax (1 + \Delta^2/\kappa^2)^{-1}$
where $\Delta =\Dpc- \Delta_N(\nbar)$ is the detuning of the probe
from the atoms-shifted cavity resonance, and $\nmax$ characterizes
the incident probe power.

\begin{figure}
\begin{center}
\includegraphics[angle = 0, width =.45\textwidth]{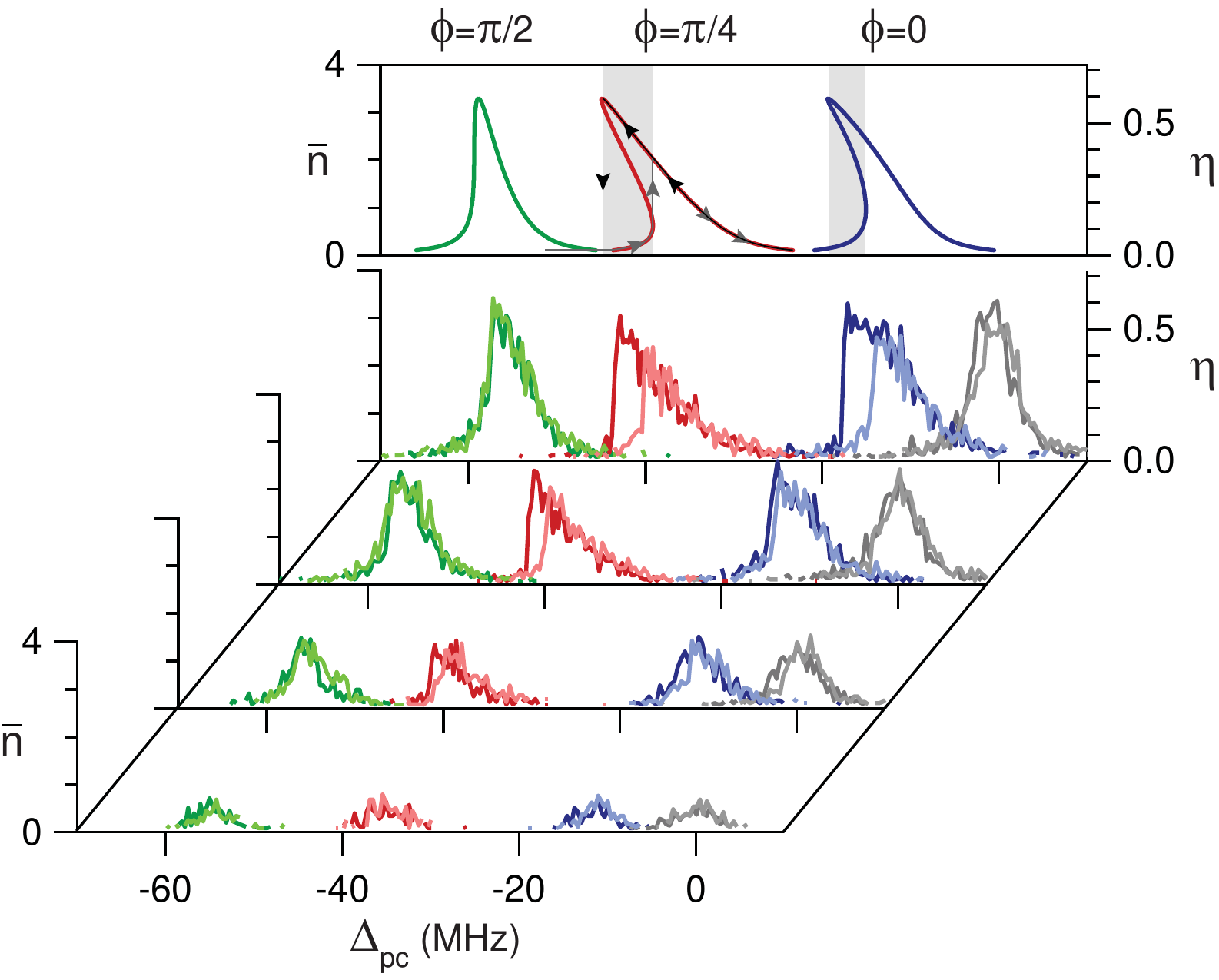}
\end{center}
\caption{(Color) Intracavity photon number $\nbar$ (or ratio of
probe/trap curvatures $\eta$) for probe light swept with positive
(lighter lines) or negative (darker lines) chirps.  Measurements for
four probe powers are shown, with $\phi_0 = 0$ (blue), $\pi/4$ (red)
or $\pi/2$ (green).  The horizontal scales are adjusted to account
for a 2\% loss of atoms between frequency sweeps.  Gray traces show
measurements for an empty cavity. Pronounced cavity nonlinearity is
observed at $\nbar \simeq 1$ (input power around 5 pW).  Here
$N=5400$, $\Dca/2\pi = -14$ GHz, and $\omz/2\pi = 32$ kHz. Top graph
shows self-consistent solutions for $\hbar$ calculated from known
experimental parameters. Gray bars show regions of bistability in
which two stable cavity and cantilever states are accessible at the
same probe power and frequency.  The expected line shape at
$\phi=\pi/2$ for probes with positive and negative chirps are
indicated.} \label{fig:lineshapes}
\end{figure}

The cavity nonlinearity is characterized using two approaches. In
the first, after preparing the atomic cantilever, we measure the
cavity transmission of a constant-power probe as its frequency was
swept slowly back and forth across the cavity resonance. The cavity
transmission lineshapes show characteristic signs of cavity
nonlinearity: displacement from the low-power cavity resonance,
asymmetry, and, for sufficiently intense probe light, hysteresis
signifying the mechanical bistability of the compressible atomic
cantilever (Fig. \ref{fig:lineshapes}).

As expected in the Lamb-Dicke regime ($k_p \sigma \ll 1$), as the
probe power is raised, the onset of cavity nonlinearity occurs first
for atoms at the location of maximum linear optomechanical coupling
($\phi_0 = \pi/4$).  For atoms centered at the node and antinode of
the probe field, residual linear coupling  \cite{note:residual}
provides a weaker nonlinear optical response at low probe powers. At
higher probe powers, quadratic optomechanical coupling dominates and
leads to a novel form of optomechanical cavity nonlinearity stemming
from variation in the strain of the compressible cantilever.   The
strength of this strain-based nonlinearity depends on the ratio
$\eta$ of the maximum curvatures of the probe-induced and mechanical
potentials acting on the atoms, given as $\eta = 4 \nbar g_0^2
\omega_r / (\Dca \omz^2)$ where $\omr = \hbar k_p^2/2 m$ is the
recoil frequency.  For $\Dca<0$, the optical forces due to the probe
field at $\phi_0=0$ subtract from the mechanical confinement of the
trapping light, causing large shifts of the cavity resonance as the
ensemble widens away from the field node; at $\eta = 1$, the
confinement to the field node is eliminated altogether.  In
contrast, for atoms positioned at the probe-field antinode
($\phi_0=\pi/2$), additional compression imparts only slight
additional shifts to the resonance, reducing the nonlinear response.

\begin{figure}[tb]
\begin{center}
\includegraphics[angle = 0, width =.4 \textwidth]{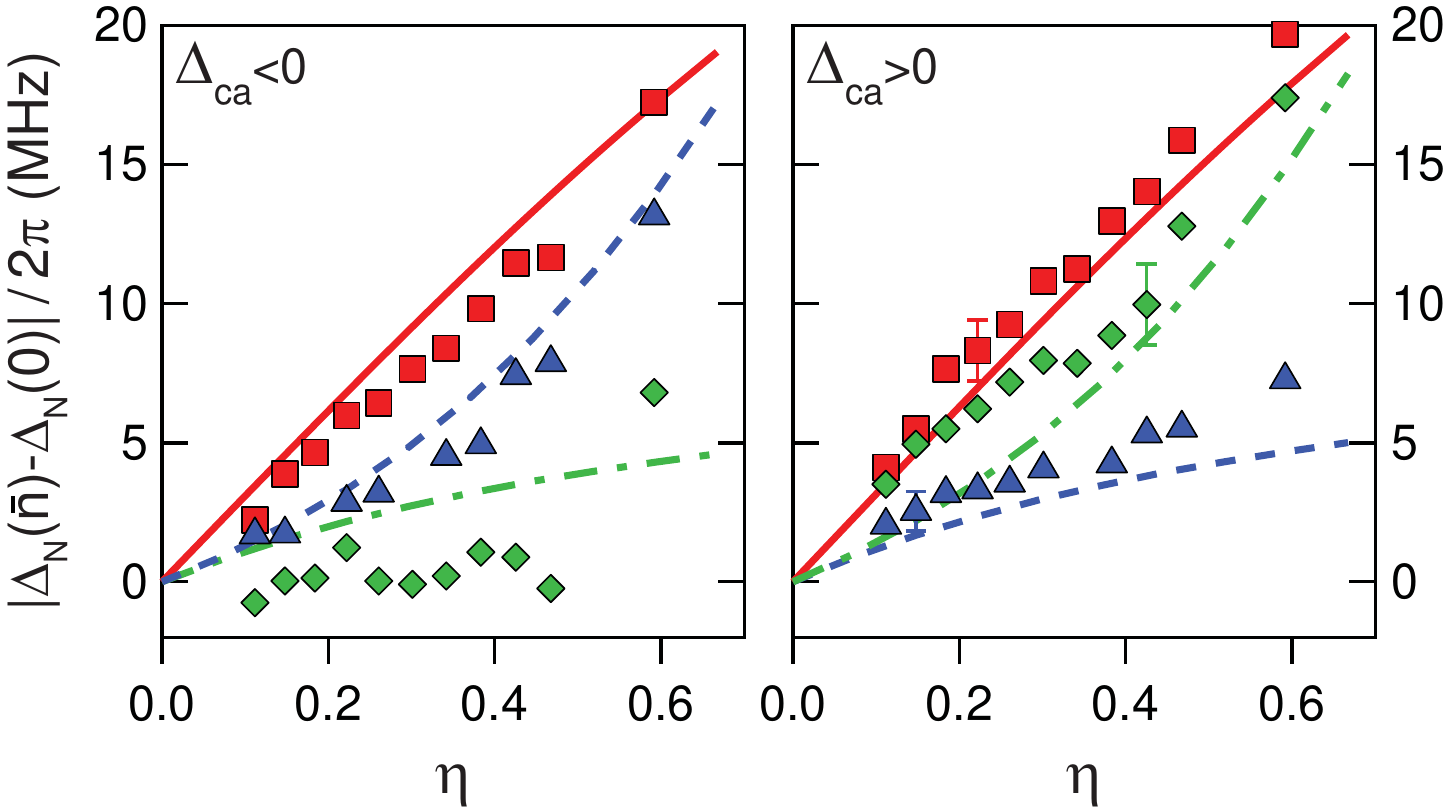}
\end{center}
\caption{(Color) Probe-induced shift of the cavity resonance vs.\
the probe/trap curvature ratio $\eta$, measured for $\phi_0=0$ (blue
triangles, dashed line), $\phi_0=\pi/4$ (red squares, solid line)
and $\phi_0=\pi/2$ (green diamonds, dot-dashed line). Representative
statistical error bars are shown.  The behaviours at the node and
antinode are interchanged as the probe potential changes from
attractive ($\Dca/2\pi = - 8 \, \mbox{GHz}$) to repulsive
($\Dca/2\pi = + 8 \, \mbox{GHz}$). The systematic discrepancy
between measurements and theory (lines) arises from atom loss during
the measurement; such heating and error are minimized at
$\phi_0=0$.} \label{fig:bve}
\end{figure}

A more quantitative measure of the cavity nonlinearity was obtained
by a second technique, in which $\nbar$ was stabilized by feedback
to the frequency of the cavity probe, fixing this frequency at a
known amount from $\Delta_N(\bar{n})$.  The cavity resonance shifts
measured in this manner (Fig.\ \ref{fig:bve}) are in good agreement
with calculations that account for the measured distribution of
atoms among lattice sites.

We now turn to measurements of a dynamical optomechanical effect,
the optomechanical frequency shift.  We distinguish between two
types of frequency shift. Considering again atoms trapped within a
single lattice site, a ``dynamic'' frequency shift arises dominantly
from linear optomechanical coupling. In a cavity probed away from
the cavity resonance, the mean intracavity photon number varies
linearly for small displacements of $\Zcm$, yielding an additional
``optical spring'' \cite{shea04spring,corb06spring} with spring
constant $K_d =  [2 F^2 \nbar \Delta/(\Delta^2 + \kappa^2)]
\sin\left(2 \phi_0\right)$.  An additional ``static'' frequency
shift arises from the quadratic optomechanical coupling,
representing the added trap curvature from a constant-intensity
intracavity probe field and quantified by the spring constant $K_s =
2 k_p F \nbar \cos\left(2 \phi_0\right)$.

We note that the static frequency shift applies to the motion of
each individual atom in the ensemble, and therefore to all axial
mechanical modes, including the linearly coupled mode. In contrast,
the latter mode is uniquely influenced by the dynamic optomechanical
frequency shift. This contrast allows us to distinguish
experimentally between the two types of shift.

\begin{figure}
\begin{center}
\includegraphics[angle = 0, width =.35\textwidth]{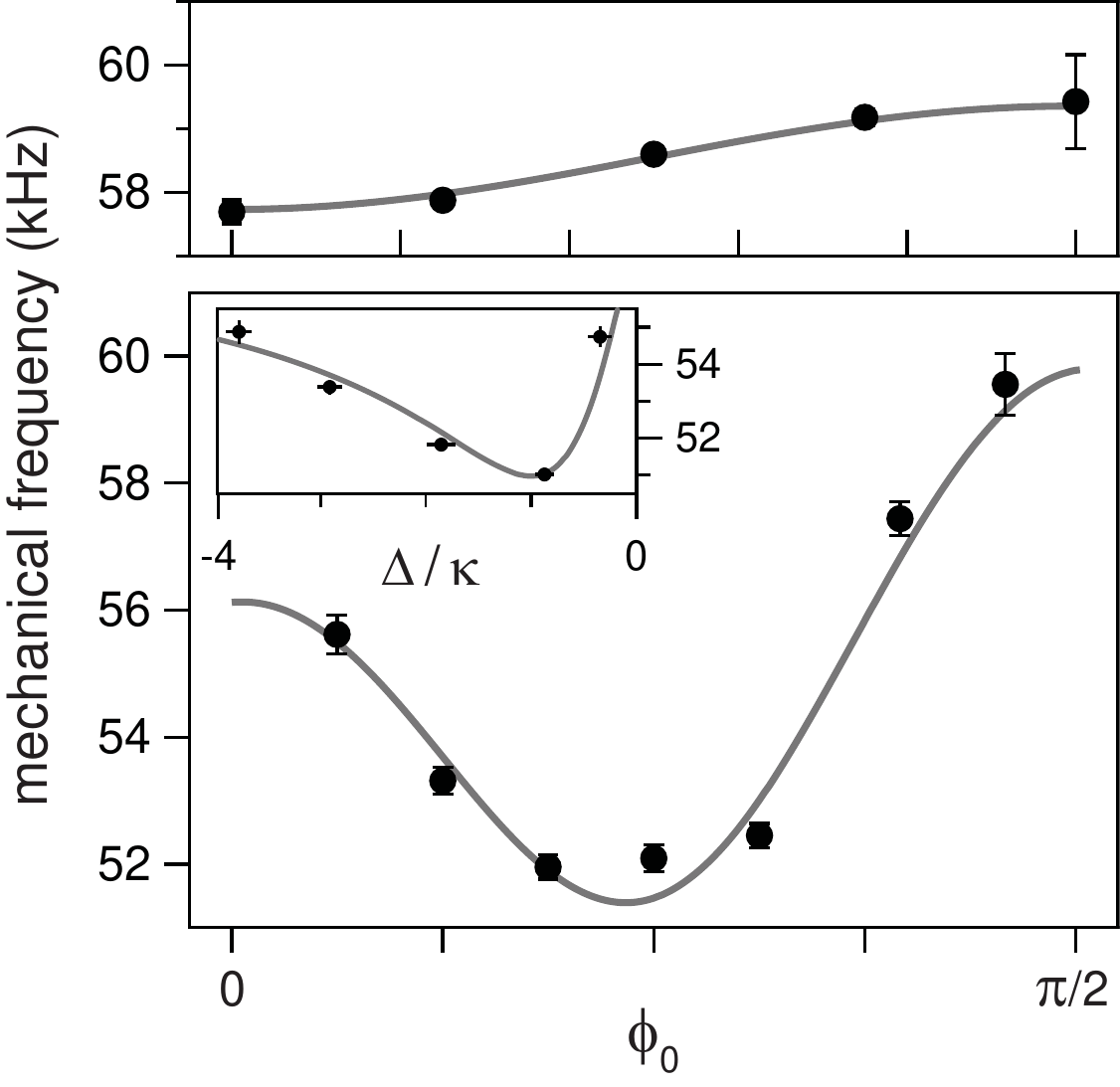}
\end{center}
\caption{Optomechanical frequency shifts.  Top: the mechanical
oscillation frequency determined via parametric heating reveals the
static frequency shift from quadratic optomechanical coupling; here,
$\Dca/2\pi\!=\!20.1 \, \mbox{GHz}$, $\omz/2\pi\!=\!58.5 \,
\mbox{kHz}$ and $\bar{n}\!=\! 0.8$. Bottom: the oscillation
frequency for the linearly coupled mode is derived from the cavity
transmission intensity spectrum; here, $\Dca/2\pi \!=\! 40 \,
\mbox{GHz}$, $\omz/2\pi\! =\! 58.5 \, \mbox{kHz}$, $\nbar \!=\! 3.5$
and $N\! =\! 3750$. Inset shows this frequency at $\phi_0=\pi/4$ for
varying detuning $\Delta$ from the cavity resonance.  Solid lines
show predictions calculated with no free parameters.}
\label{fig:shifts}
\end{figure}

To observe the static frequency shift, we measured the resonant
atomic oscillation frequency via parametric heating of the ensemble
(Fig.\ \ref{fig:shifts}). With probe light stabilized at fixed
$\nbar$ and $\Delta$, we modulated the trap laser intensity at fixed
frequency and then measured the fraction of atoms heated out of the
optical trap by absorption imaging. Such modulation drives all
mechanical modes, the influence of the linearly coupled mode being
negligible with large $N$. Therefore, the dominant atom loss is
observed at the frequency $2 \sqrt{\omz^2 + K_s/(N m)}$.

Alternately, we selectively detected the linearly coupled mode by
monitoring temporal variations of the light transmitted through the
cavity. Here, this mechanical mode is driven by technical and
quantum noise of the probe light intensity \cite{note:servo}.  Our
optomechanical system amplifies this noise with a frequency
dependent gain \cite{manc94,fabr94noise,mari10signature,verl10} that
is visible in the power spectrum of the transmitted probe light.
From a fit to this gain spectrum, we extracted the resonance
frequency $\sqrt{\omz^2 + (K_s + K_d)/(N m)}$ of the linearly
coupled mode. The largest optomechanical frequency shift is observed
at positions with maximal linear coupling ($\phi_0 = \pi/4$) and at
$\Delta = - \kappa$. The quantitative agreement with theoretical
predictions with no free parameters is remarkable.

Our tunable cavity optomechanical system lies deeply in the quantum
regime, as the cantilever is prepared near its ground state and its
motion is dominated by quantum radiation pressure fluctuations. This
capability should enable future studies of quantum optomechanical
effects and of the role of linear and quadratic coupling in such
phenomena. This platform also offers the possibility to enter the
single-photon strong optomechanical coupling regime
\cite{murc08backaction,ludw08instability}, in which the nature of
optomechanical effects remains poorly understood.

This work was supported by the NSF and the AFOSR.  T.B. acknowledges
support from Le Fonds Qu\'{e}b\'{e}cois de la Recherche sur la
Nature et les Technologies, and D.M.S.-K. from the Miller Institute
for Basic Research in Science.


\begin{thebibliography}{10}
\bibitem{note:transverse}%
  \BibitemOpen
  \bibinfo {note} {The transverse extent of the ensemble only varies the
  effective cavity-atom coupling $g_0$, and not the contrast.}%
  \bibAnnoteFile{Stop}{note:transverse}%
\bibitem{note:residual}%
  \BibitemOpen
  \bibinfo {note} {For atoms in three neighboring wells with population ratios
  1:2:1, linear coupling to the motions of atoms in the outer wells has
  relative strength $\sin^2 (2 \pi k_p/k_t) / 2 = 1/8$ compared to the maximum
  linear coupling; see Ref.\ \cite{murc08backaction}.}%
  \bibAnnoteFile{Stop}{note:residual}%
\bibitem{note:servo}%
  \BibitemOpen
  \bibinfo {note} {The small servo bandwidth of several kHz ensured that
  mechanical oscillations of the ensemble were not influenced by this
  feedback.}%
  \bibAnnoteFile{Stop}{note:servo}%
\end{thebibliography}
%
\end{document}